\documentclass[conference]{IEEEtran}
\IEEEoverridecommandlockouts
\usepackage{amsmath,amssymb,amsfonts}
\usepackage{algorithmic}
\usepackage{graphicx}
\usepackage{textcomp}
\usepackage{xcolor}
\usepackage{cite,epsfig,bm}
\usepackage{multirow}
\usepackage{booktabs}
\usepackage{float} 
\usepackage{subfigure} 
\usepackage{caption}
\usepackage{threeparttable}
\usepackage{url}
\def\BibTeX{{\rm B\kern-.05em{\sc i\kern-.025em b}\kern-.08em
    T\kern-.1667em\lower.7ex\hbox{E}\kern-.125emX}}
\begin{document}

\title{Quality Assessment of Stereoscopic 360-degree Images from Multi-viewports\\
\thanks{\textsuperscript{*} Equal contribution.}
}

\author{\IEEEauthorblockN{Jiahua Xu\textsuperscript{*}, Ziyuan Luo\textsuperscript{*}, Wei Zhou, Wenyuan Zhang and Zhibo Chen}
\IEEEauthorblockA{\textit{CAS Key Laboratory of Technology in Geo-Spatial Information Processing and Application System} \\
\textit{University of Science and Technology of China}\\
Hefei 230027, China \\
chenzhibo@ustc.edu.cn}
}

\maketitle

\begin{abstract}
Objective quality assessment of stereoscopic panoramic images becomes a challenging problem owing to the rapid growth of 360-degree contents. Different from traditional 2D image quality assessment (IQA), more complex aspects are involved in 3D omnidirectional IQA, especially unlimited field of view (FoV) and extra depth perception, which brings difficulty to evaluate the quality of experience (QoE) of 3D omnidirectional images. In this paper, we propose a multi-viewport based full-reference stereo 360 IQA model. Due to the freely changeable viewports when browsing in the head-mounted display, our proposed approach processes the image inside FoV rather than the projected one such as equirectangular projection (ERP). In addition, since overall QoE depends on both image quality and depth perception, we utilize the features estimated by the difference map between left and right views which can reflect disparity. The depth perception features along with binocular image qualities are employed to further predict the overall QoE of 3D 360 images. The experimental results on our public Stereoscopic OmnidirectionaL Image quality assessment Database (SOLID) show that the proposed method achieves a significant improvement over some well-known IQA metrics and can accurately reflect the overall QoE of perceived images. 
\end{abstract}


\section{Introduction}
Immersive media data such as stereoscopic omnidirectional images and videos suffer from diverse quality degradations ranging from acquisition, compression, transmission to display \cite{chen2018blind}, thus it is of great importance for automatically predicting the perceptual quality of 3D 360-degree contents to optimize the coding and processing technologies and maximize the user quality of experience (QoE) \cite{ebrahimi2009quality,perrin2015multimodal}. Compared with conventional 2D image quality assessment (IQA), it is more challenging to evaluate the quality of stereoscopic panoramic images due to the unlimited field of view (FoV) and extra dimension of depth perception \cite{xu2018subjective}. Although IQA has been researched in recent years \cite{sheikh2006statistical,zhai2015dual,nizami2019distortion}, a few works have been done to predict the perceptual quality of stereo 360 images which remains an intractable research problem.

Image quality assessment is mainly divided into two categories: subjective IQA and objective IQA \cite{seshadrinathan2010study} and it is the same for stereoscopic omnidirectional image quality assessment (SOIQA) \cite{zhang2017subjective}. Though subjective SOIQA can generate the mean opinion scores (MOSs) of all the subjects as the most accurate quality evaluation \cite{yamagishi2013subjective}, it is usually unpractical in real applications due to the time-consuming and labor-intensive attributes. Hence, the objective metrics designed for SOIQA are in great demand.

Up to now, several algorithms have been proposed for stereoscopic image quality assessment (SIQA) and omnidirectional  image quality assessment (OIQA). To deal with SIQA, traditional 2D IQA methods such as peak signal-to-noise ratio (PSNR), structural similarity (SSIM) \cite{wang2004image}, multi-scale structural similarity (MS-SSIM) \cite{wang2003multiscale} were performed on the left and right view images separately in the early stage \cite{yasakethu2008quality}. Later, disparity map between two views was employed to make an improvement \cite{benoit2009quality}. The models mentioned above show good performance on symmetrical distortion while their correlations with subjective scores are rather low for asymmetrical distortion. Then, binocular vision properties of the human visual system (HVS) were investigated and binocular fusion, rivalry, suppression models were widely used in 3D IQA \cite{zhou2016binocular,saygili2010quality,shao2013perceptual,lin2014quality}. 

In terms of OIQA, several PSNR based metrics including spherical PSNR (S-PSNR) \cite{yu2015framework}, weighted-to-spherically-uniform PSNR (WS-PSNR) \cite{sun2017weighted}, craster parabolic projection PSNR (CPP-PSNR) \cite{zakharchenko2016quality} were proposed by considering the characteristics of 360-degree images. They are efficient and easy to be integrated into codecs but the prediction accuracy is far from satisfactory. Then, some perception-driven IQA metrics for 360 contents were designed via machine learning \cite{xu2018assessing,yang2017objective} and deep learning \cite{kim2019deep,yang20183d}. Chen \emph{et al.} further incorporated SIQA and OIQA and developed a predictive coding based model for 3D 360 image quality assessment \cite{chen2019stereoscopic}. Compared with \cite{chen2019stereoscopic}, our proposed model not only predicts the perceptual image quality, but also estimates the overall QoE which is not mentioned previously. Note that overall QoE is a measure of the overall level of customer satisfaction with the perceived image, it considers not only image quality but also other factors such as depth perception, visual comfort, etc.

\begin{figure*}[htbp]
\vspace{-0.8cm}
  \centerline{\includegraphics[width=17cm]{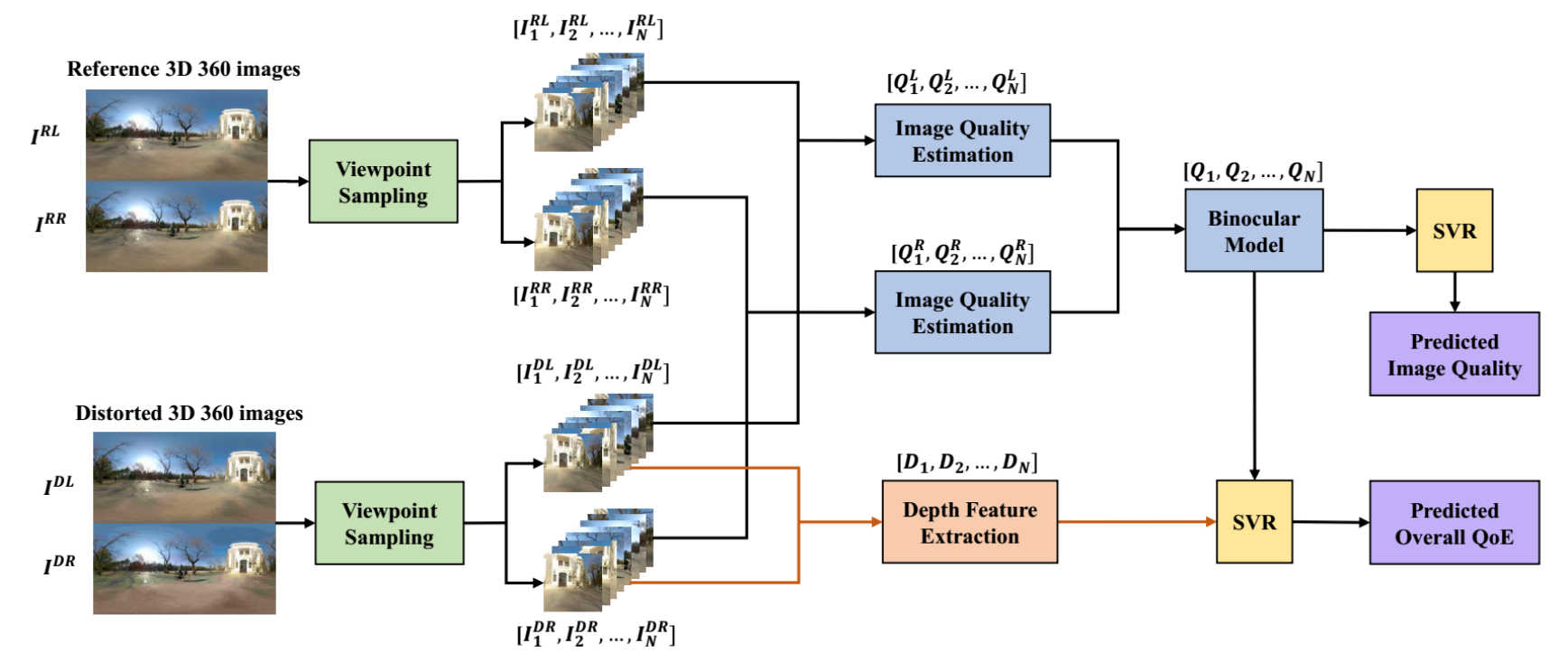}}
  \caption{The architecture of the proposed stereoscopic omnidirectional image quality assessment (SOIQA) model.}
  \centering
\label{figure1}
\vspace{-0.3cm}
\end{figure*}

In this paper, motivated by the free FoV characteristic and binocular properties of stereoscopic omnidirectional images, we propose a novel multi-viewport based algorithm for SOIQA which considers the image quality inside FoV and the HVS mechanism. Moreover, the viewpoints are specially selected instead of uniform sampling. To predict the overall QoE of 3D 360-degree images, depth perception related features are estimated from the difference map between left and right view images and further integrated with image quality scores. We test the proposed model on the self-built public stereoscopic omnidirectional image quality assessment database (SOLID) \cite{xu2018subjective} and the experimental results verify the effectiveness of this method.

The rest of the paper is organized as follows. The proposed multi-viewport based SOIQA metric is described in Section II. In Section III, we present the experimental results and then conclude in Section IV.

\section{Proposed Stereoscopic Omnidirectional Image Quality Assessment Model}

In the proposed multi-viewport based SOIQA model, it takes the reference and distorted stereo 360 image pairs as input and returns the predicted image quality and overall QoE. We first illustrate the framework of the proposed model and the detailed components are described in the following paragraphs.

\subsection{Architecture}

The framework of the proposed SOIQA model is depicted in Fig. \ref{figure1}. It is composed of the viewport sampling, binocular image quality estimation, depth feature extraction and support vector regression (SVR). At first, the reference and distorted 3D panoramic image pairs are sampled as several independent viewport images with an FoV of 90 degree. Then, we use the well-known full-reference 2D IQA metric feature similarity (FSIM) index \cite{zhang2011fsim} to predict the image quality for left viewport images and right viewport images separately. The binocular model is adopted to allocate the rivalry dominance and compute the aggerated score for stereo viewport images. In addition, we subtract the left and right viewport images to get the difference map and extract the depth perception related features from it. Finally, the viewport image quality features as well as the depth perception features are regressed onto the final perceptual image quality and overall QoE.

\subsection{Viewpoint Sampling}
\begin{figure}[hbtp]
	\vspace{-0.3cm}
    \centering
    \subfigure[]{
        \includegraphics[height=2.8cm]{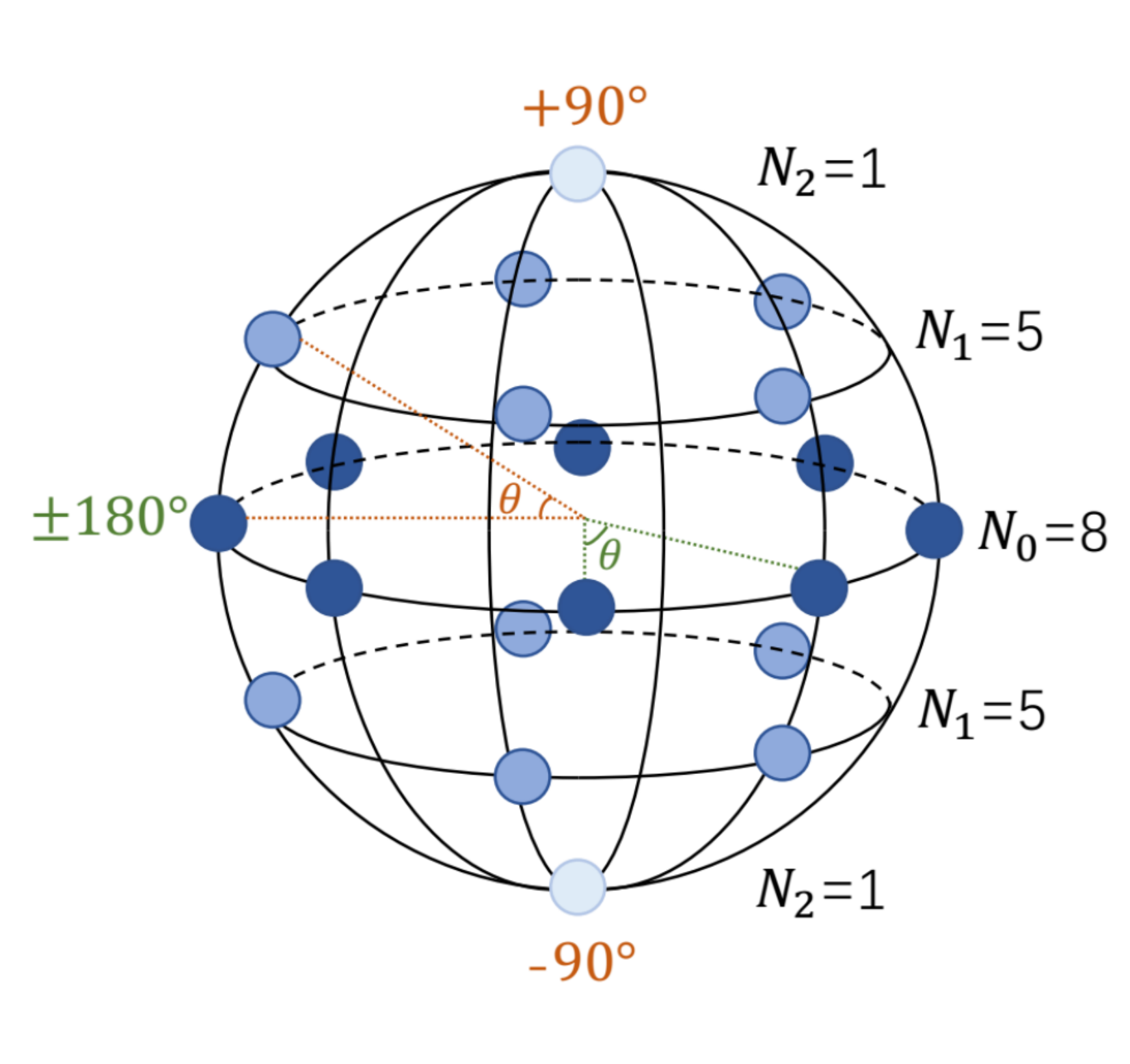}
    }
    \subfigure[]{
        \includegraphics[height=2.8cm]{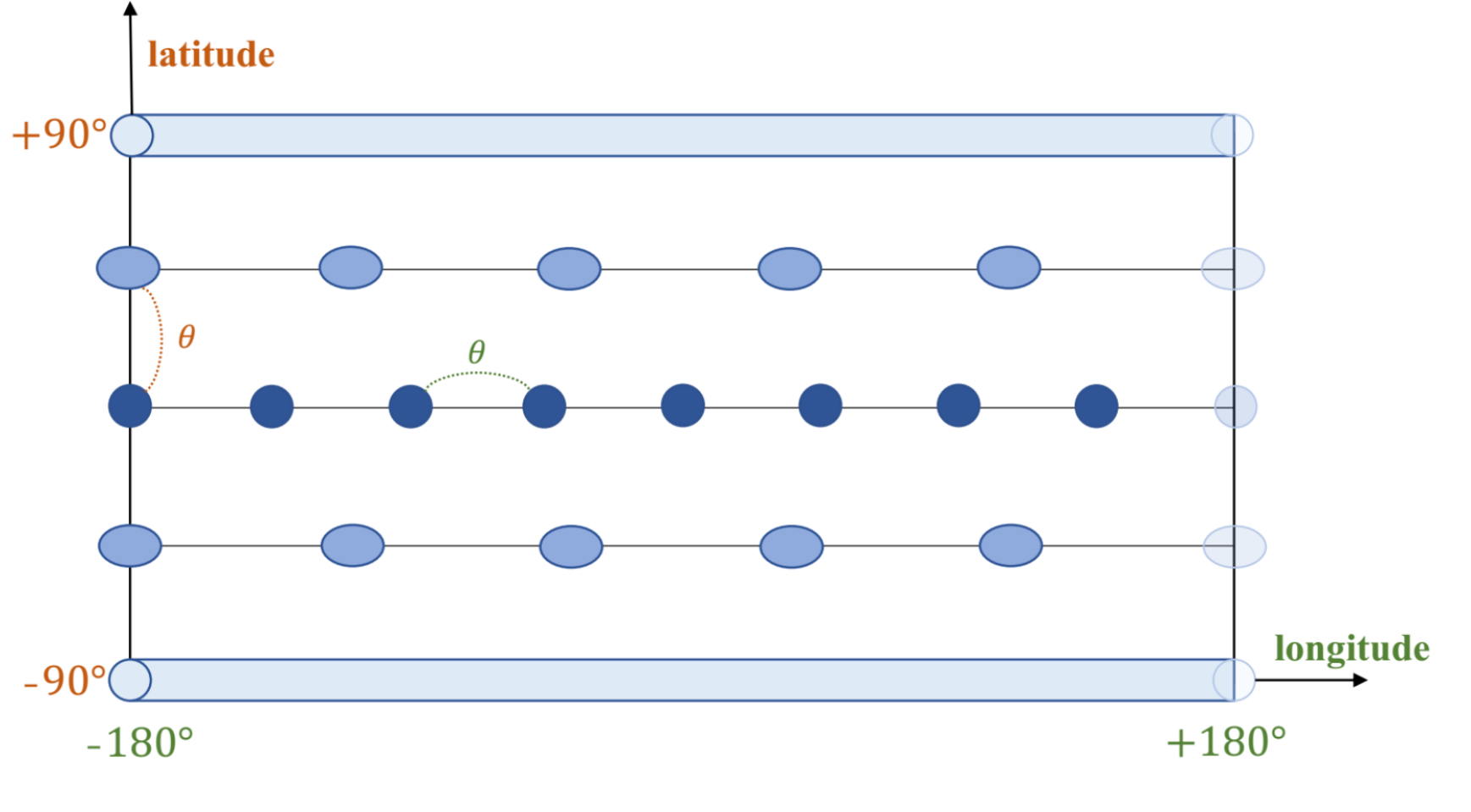}
    }  
    \caption{An example of sampling viewpoints when $N_{0}=8$ and $\theta=45^{\circ}$, (a) sampling on the sphere, (b) sampling on the plane.}
    \label{figure2}
    \vspace{-0.3cm} 
\end{figure}

Omnidirectional images are viewed in the sphere surface while transmitted and stored in the 2D format. As a result, evaluating the 2D format omnidirectional image is different from the actual viewing experience. Moreover, 360 images rendered in the equirectangular projection (ERP) format usually stretch polar regions and generate projection deformation. To solve the above problems, we design a novel viewport selection strategy instead of uniform sampling on the ERP format. Assume $N_0$ viewpoints are equidistantly sampled on the equator, the other points are chosen as follows:
\begin{equation}\label{1}
\theta {\rm{ = }}\frac{{360^{\circ}}}{{{N_0}}},
\end{equation}
\begin{equation}\label{2}
{N_1}{\rm{ = }}\left\lfloor {{N_0}\cos \theta } \right\rfloor,{N_2}{\rm{ = }}\left\lfloor {{N_0}\cos 2\theta } \right\rfloor,\,...\,,
\end{equation}
where $N_1$ and $N_2$ represent the number of points sampled on $\theta$ and $2\theta$ degrees north or south latitude. The sampling procedure lasts until the maximum latitude reaches 90 degree. These viewpoints are uniformly distributed on the particular latitudes. Fig. \ref{figure2} gives an example when $N_0=8$ and $\theta=45^{\circ}$. Note that the viewpoints only sampled once at the south and north poles.

\subsection{Image Quality Estimation}

In this module, we process the 3D viewport images as traditional stereoscopic images covering the $90^{\circ}$ FoV range and compute the perceptual viewport image quality according to the binocular rivalry model \cite{wang2015quality}. For stereo image pairs, the left and right view images tend to share different weights which are related to binocular energies. Therefore, we adopt the local variances to compute the energy maps of both views as done in\cite{wang2006spatial}. Then, the local energy ratio maps $\bm{R}^{L}_{n}$ and $\bm{R}^{R}_{n}$ of the $n$-th left and right viewport images can be denoted as:
\begin{equation}\label{3}
\bm{R}_{n}^L = \frac{\bm{E}_{n}^{DL}}{\bm{E}_{n}^{RL}} \quad and \quad  \bm{R}_{n}^R = \frac{\bm{E}_{n}^{DR}}{\bm{E}_{n}^{RR}} ,
\end{equation}
where $\bm{E}^{DL}_{n}$, $\bm{E}^{RL}_{n}$, $\bm{E}^{DR}_{n}$ and $\bm{E}^{RR}_{n}$ indicate the energy maps of distorted and reference images for the $n$-th viewports. Since the HVS prefer high-energy regions which involve more information and are easier to attract visual attention, we employ the energy weighted pooling method \cite{wang2015quality} to stress significance on high-energy image regions in binocular rivalry as follows:
\begin{equation}\label{4.1}
g_n^L = \frac{{\sum_{i,j} {\bm{E}_n^{DL}(i,j)\bm{R}_n^L(i,j)} }}{{\sum_{i,j} {\bm{E}_n^{DL}(i,j)} }} ,
\end{equation}
\begin{equation}\label{4.2}
g_n^R = \frac{{\sum_{i,j} {\bm{E}_n^{DR}(i,j)\bm{R}_n^R(i,j)} }}{{\sum_{i,j} {\bm{E}_n^{DR}(i,j)} }},
\end{equation}
where $g^{L}_{n}$ and $g^{R}_{n}$ are calculated by summation on the full energy and ratio maps. Hence, they denote the level of dominance for the $n$-th left and right viewport images. After that, we compute the weights for left and right views as follows:
\begin{equation}\label{5}
w_{n}^L = \frac{{g{{_{n}^L}^2}}}{{g{{_{n}^L}^2} + g{{_{n}^R}^2}}} \quad and \quad w_{n}^R = \frac{{g{{_{n}^R}^2}}}{{g{{_{n}^L}^2} + g{{_{n}^R}^2}}}.
\end{equation}

Finally, the binocular image quality (feature similarity) $Q_{n}$ for the $n$-th viewport image is calculated by a weighted average of both views:
\begin{equation}\label{6}
{Q_{n}} = w_{n}^LQ_{n}^L + w_{n}^RQ_{n}^R,
\end{equation}
\begin{equation}\label{7}
Q_{n}^L = FSIM(\bm{I}_n^{RL},\bm{I}_n^{DL}) \ and \ Q_{n}^R = FSIM(\bm{I}_n^{RR},\bm{I}_n^{DR}),
\end{equation}
where image quality $Q^{L}_{n}$ and $Q^{R}_{n}$ for left and right view images are obtained through FSIM \cite{zhang2011fsim} which can accurately predict the quality of  2D images. $\bm{I}_n^{RL}$, $\bm{I}_n^{DL}$, $\bm{I}_n^{RR}$ and $\bm{I}_n^{DR}$ represent the $n$-th reference and distorted viewport images.

\subsection{Depth Feature Extraction}
\begin{figure}[hbtp]
	\centering
	\subfigure[]{
		\includegraphics[height=2cm]{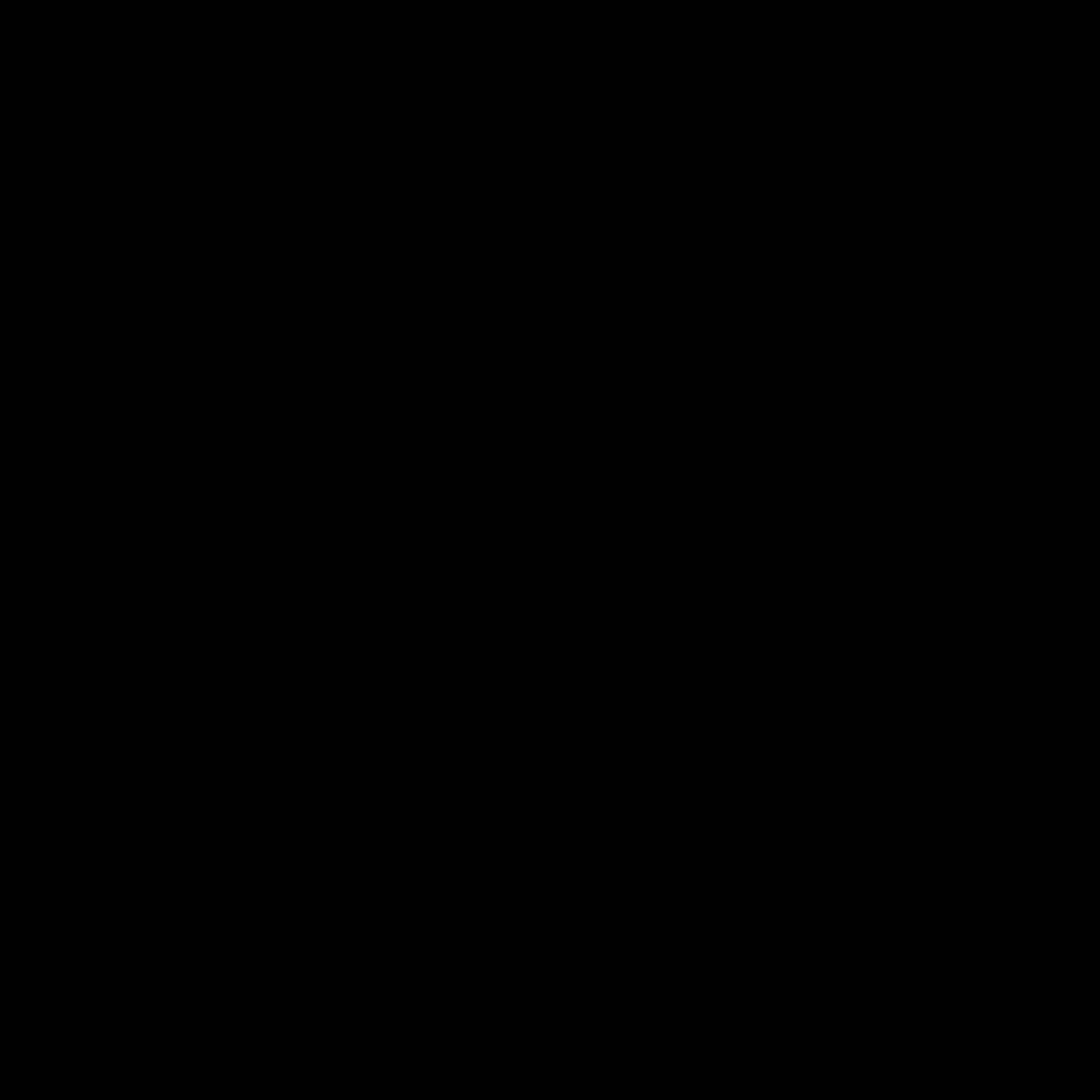}
	}
	\subfigure[]{
		\includegraphics[height=2cm]{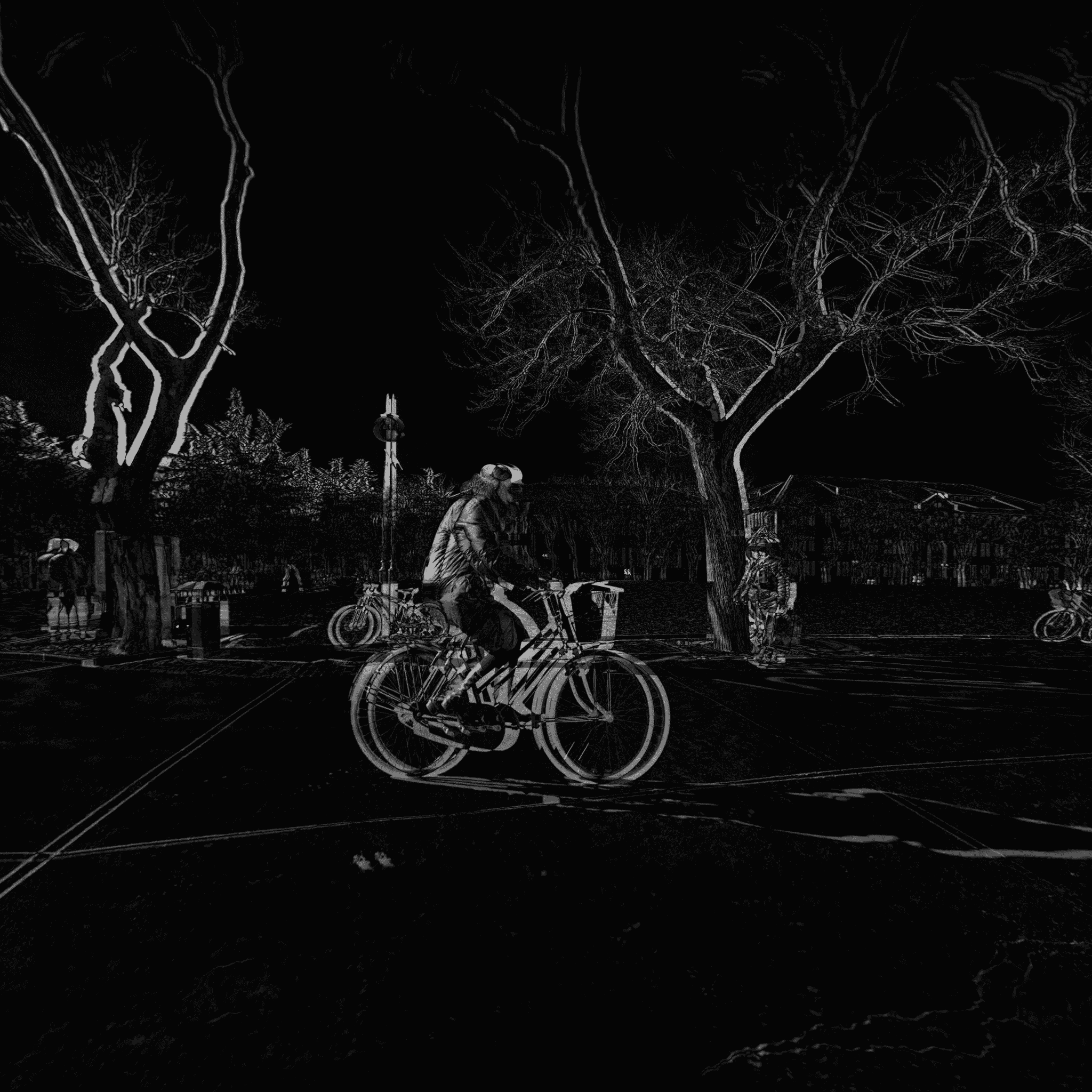}
	}
	\subfigure[]{
		\includegraphics[height=2cm]{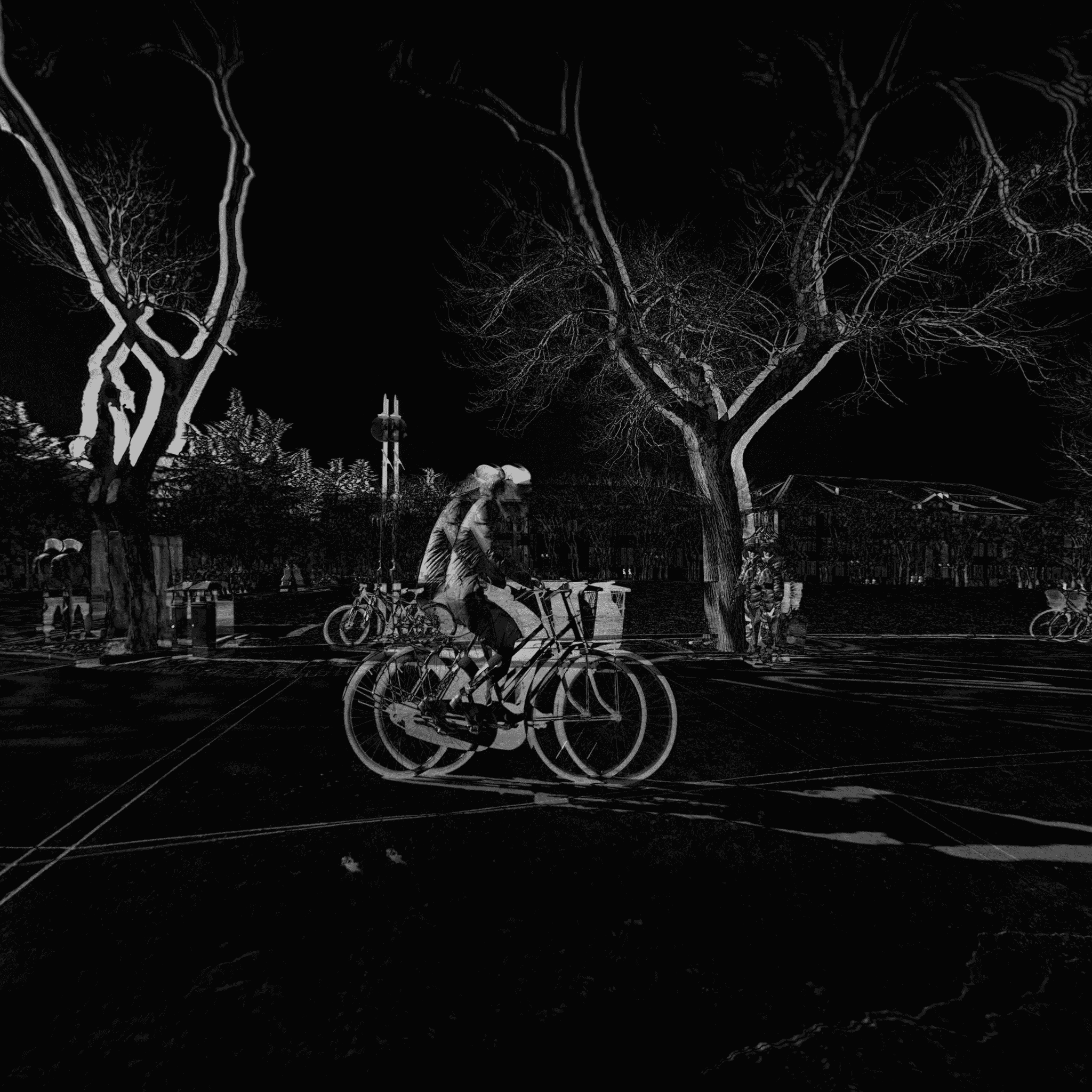}
	}  
	\caption{The difference map for viewport images with (a) zero disparity, (b) medium disparity, (c) large disparity.}
	\label{figure3}
\end{figure}

As analyzed in \cite{chen2019stereoscopic}, depth perception is dominated by disparity, so we subtract the left and right viewport images to show the discrepancy between them. To some extent, the difference map also reflects the disparity which is illustrated in Fig. \ref{figure3}. As we can observe from Fig. \ref{figure3}, the difference map of zero disparity is a totally black image. When the disparity is larger, more information is contained in the difference map. Thus, we compute the entropy of the difference map for the $n$-th viewport as a depth perception related feature $D_n$ as follows:
\begin{equation}\label{7}
\bm{S}_n^ -  = \left|\bm{I}_n^{DL} - \bm{I}_n^{DR}\right|
\end{equation}
\begin{equation}\label{8}
{D_n} = {\rm{ - }}\sum\limits_i {{p_i}\log {p_i}}
\end{equation}
where $\bm{I}^{DL}_n$ and $\bm{I}^{DR}_n$ represent the grayscale distorted stereo pairs for the $n$-th viewport since humans are more sensitive to luminance. $p_i$ denotes the probability of $i$-th gray level appearing in the difference map $\bm{S}^-_n$. Note that depth feature extraction is only performed in the distorted images because distortion has little effect on the depth perception and the distorted images are actually observed by the subjects \cite{xu2018subjective}.

\subsection{Quality Regression}
After computing binocular feature similarities and extracting depth features from $N$ viewport images, SVR is adopted to map them onto the final perceptual image quality and overall QoE. Note that the LibSVM package is utilized in our experiment to implement the SVR with a radial basis function kernel.

\section{Experimental Results and Analysis}
In this section, we test the proposed model on the self-built public database SOLID \cite{xu2018subjective}. The experimental results demonstrate the effectiveness of our algorithm and it outperforms several classic IQA metrics. In our experiment settings, $N_0$ equals 8 which means 20 viewpoints are selected as shown in Fig. \ref{figure2}. Moreover, each viewport image covers a $90^{\circ}$ FoV to avoid heavy projection deformation.

\subsection{Database and Performance Measure}

\begin{figure}[htbp]
	\centering
	\subfigure[]{
		\includegraphics[width=1.1cm]{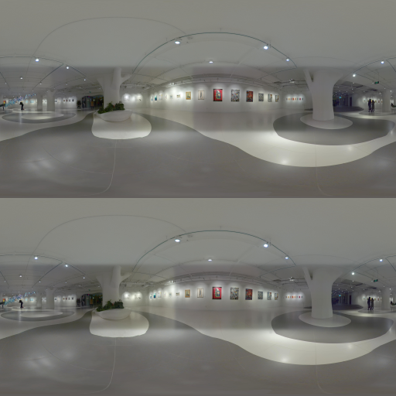}
	}
	\subfigure[]{
		\includegraphics[width=1.1cm]{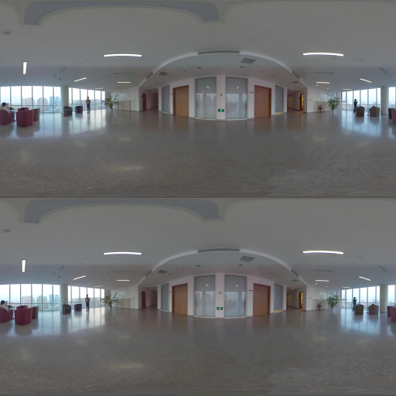}
	}
	\subfigure[]{
		\includegraphics[width=1.1cm]{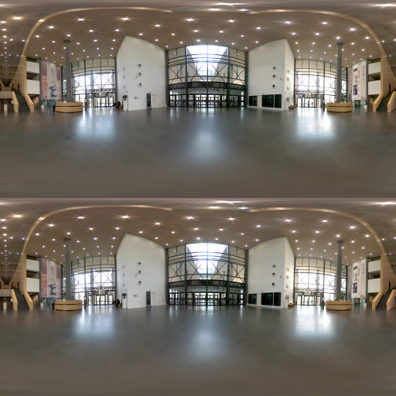}
	}
	\subfigure[]{
		\includegraphics[width=1.1cm]{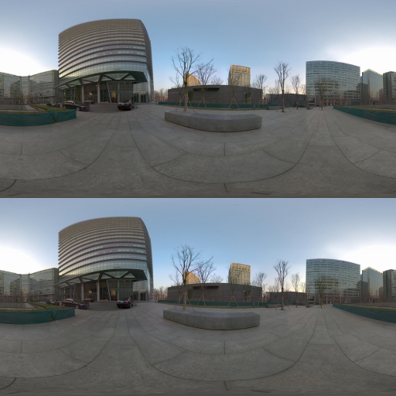}
	}
	\subfigure[]{
		\includegraphics[width=1.1cm]{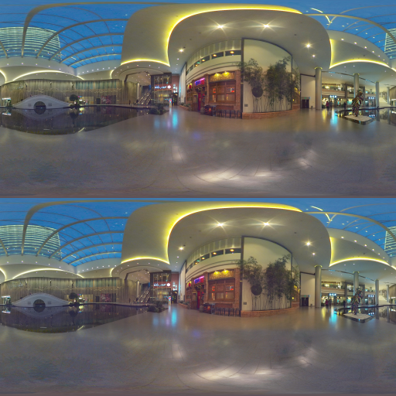}
	}
	\subfigure[]{
		\includegraphics[width=1.1cm]{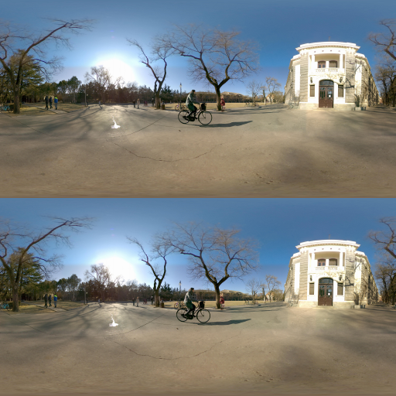}
	}
	\caption{Reference images in SOLID database. }
	\label{figure4}
	\vspace{-0.5cm} 
\end{figure}

To our best knowledge, the SOLID database is the only publicly available database built for stereoscopic omnidirectional image quality assessment which consists of 84 symmetrically and 192 asymmetrically distorted images. They originate from 6 reference images with two distortion types (JPEG and BPG compression) and three depth levels (zero, medium and large disparity). The detail configrations can be found in the webpage $\footnote{\url{http://staff.ustc.edu.cn/~chenzhibo/resources.html}}$. The reference images in the SOLID database are shown in Fig. \ref{figure4}. Each pristine and degraded image are equipped with the MOS values of image quality, depth perception and overall QoE in the range of 1 to 5, where higher subjective score means better quality.

Three commonly used criteria are utilized for performance evaluation in our experiment, namely Spearman's rank order correlation coefficient (SROCC), Pearson's linear correlation coefficient (PLCC) and root mean squared error (RMSE). SROCC measures prediction monotonicity while PLCC and RMSE measure the prediction accuracy. Higher SROCC, PLCC and lower RMSE indicate good correlation with subjective judgments. Before computing PLCC and RMSE, a five-parameter logistic function is applied to maximize the correlations between objective metrics and subjective scores \cite{sheikh2006statistical }.

\subsection{Performance Comparison}
There are six reference images in the SOLID database, we randomly split the database into 67\% training and 33\% testing set according to the reference content as done in \cite{kim2019deep}. The cross validation is performed 1000 times and we calculate the median SROCC, PLCC and RMSE as performance measurement.

\begin{table}[htbp]
\centering
\caption{\textsc{Performance Evaluation on the SOLID Database \cite{xu2018subjective}.}}
\label{table1}
\scalebox{0.82}{
\begin{tabular}{@{}c|c|ccc|ccc@{}}
\toprule
 &  & \multicolumn{3}{c|}{Image Quality} & \multicolumn{3}{c}{Overall QoE} \\ \midrule
Type & Metric & PLCC & SROCC & RMSE & PLCC & SROCC & RMSE \\ \midrule
\multirow{4}{*}{2D IQA} & PSNR & 0.629 & 0.603 & 0.789 & 0.546 & 0.506 & 0.696 \\
 & SSIM \cite{wang2004image} & 0.882 & 0.888 & 0.478 & 0.738 & 0.748 & 0.561 \\
 & MS-SSIM \cite{wang2003multiscale}& 0.773 & 0.755 & 0.643 & 0.645 & 0.620 & 0.635 \\
 & FSIM \cite{zhang2011fsim}& 0.889 & 0.883 & 0.465 & 0.747 & 0.748 & 0.552 \\ \midrule
\multirow{3}{*}{2D OIQA} & S-PSNR \cite{yu2015framework}& 0.593 & 0.567 & 0.816 & 0.507 & 0.475 & 0.716 \\
 & WS-PSNR \cite{sun2017weighted}& 0.585 & 0.559 & 0.823 & 0.499 & 0.470 & 0.720 \\
 & CPP-PSNR \cite{zakharchenko2016quality}& 0.593 & 0.566 & 0.817 & 0.506 & 0.475 & 0.716 \\ \midrule
\multirow{3}{*}{3D IQA} & Chen \cite{chen2013full}& 0.853 & 0.827 & 0.530 & 0.661 & 0.636 & 0.623 \\
 & W-SSIM \cite{wang2015quality}& 0.893 & 0.891 & 0.457 & 0.743 & 0.748 & 0.556 \\
 & W-FSIM \cite{wang2015quality} & 0.889 & 0.885 & 0.464 & 0.746 & 0.750 & 0.553 \\ \midrule
\multirow{2}{*}{3D OIQA} & SOIQE \cite{chen2019stereoscopic} & 0.927 & 0.924 & 0.383 & 0.803 & 0.805 & 0.495 \\
 & Proposed & \textbf{0.939} & \textbf{0.928} & \textbf{0.351} & \textbf{0.935} & \textbf{0.925} & \textbf{0.294} \\ \bottomrule
\end{tabular}}
\end{table}

\begin{figure}[hbtp]
	\centering
	\subfigure[]{
		\includegraphics[height=3cm]{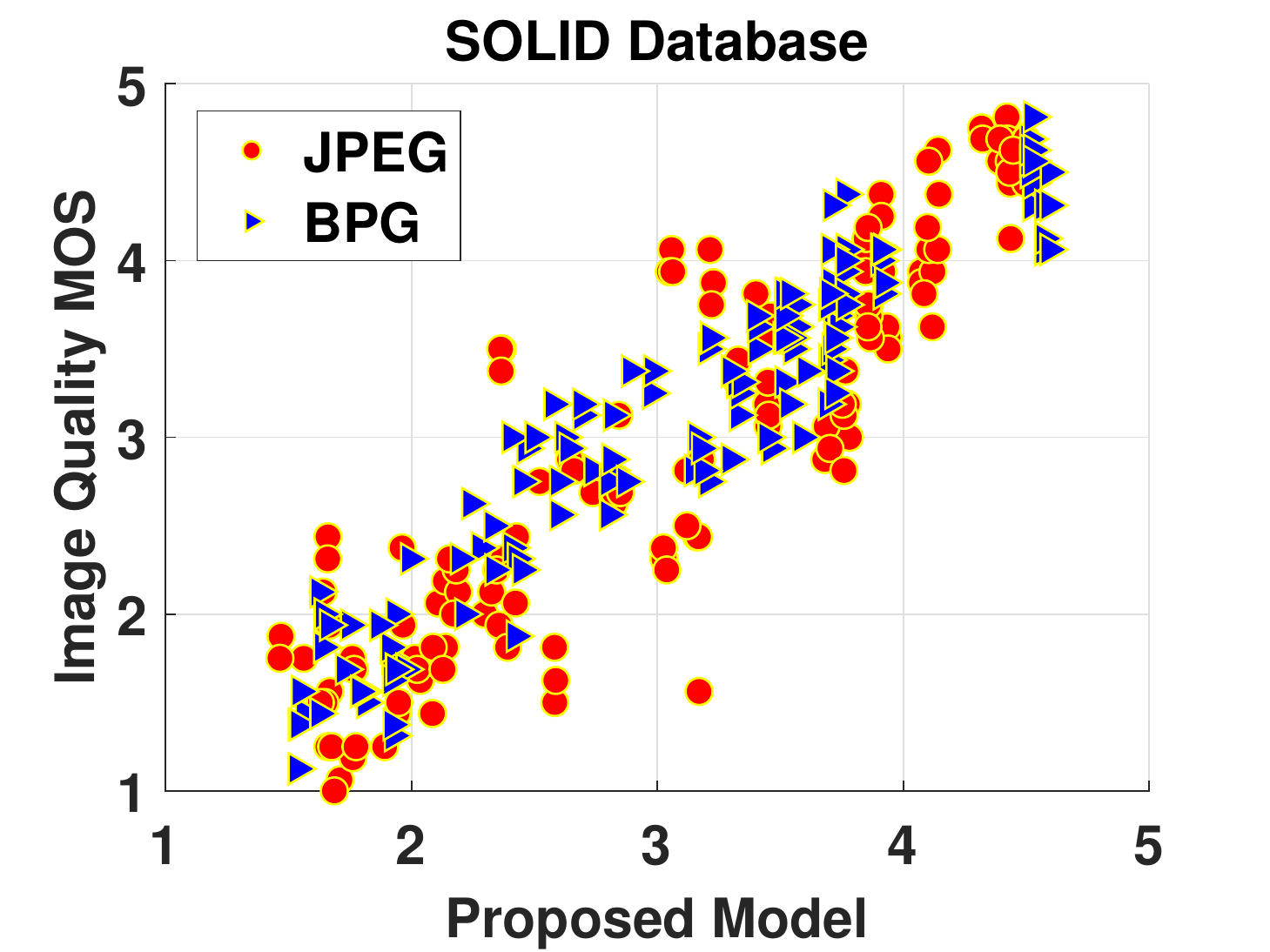}
	}
	\subfigure[]{
		\includegraphics[height=3cm]{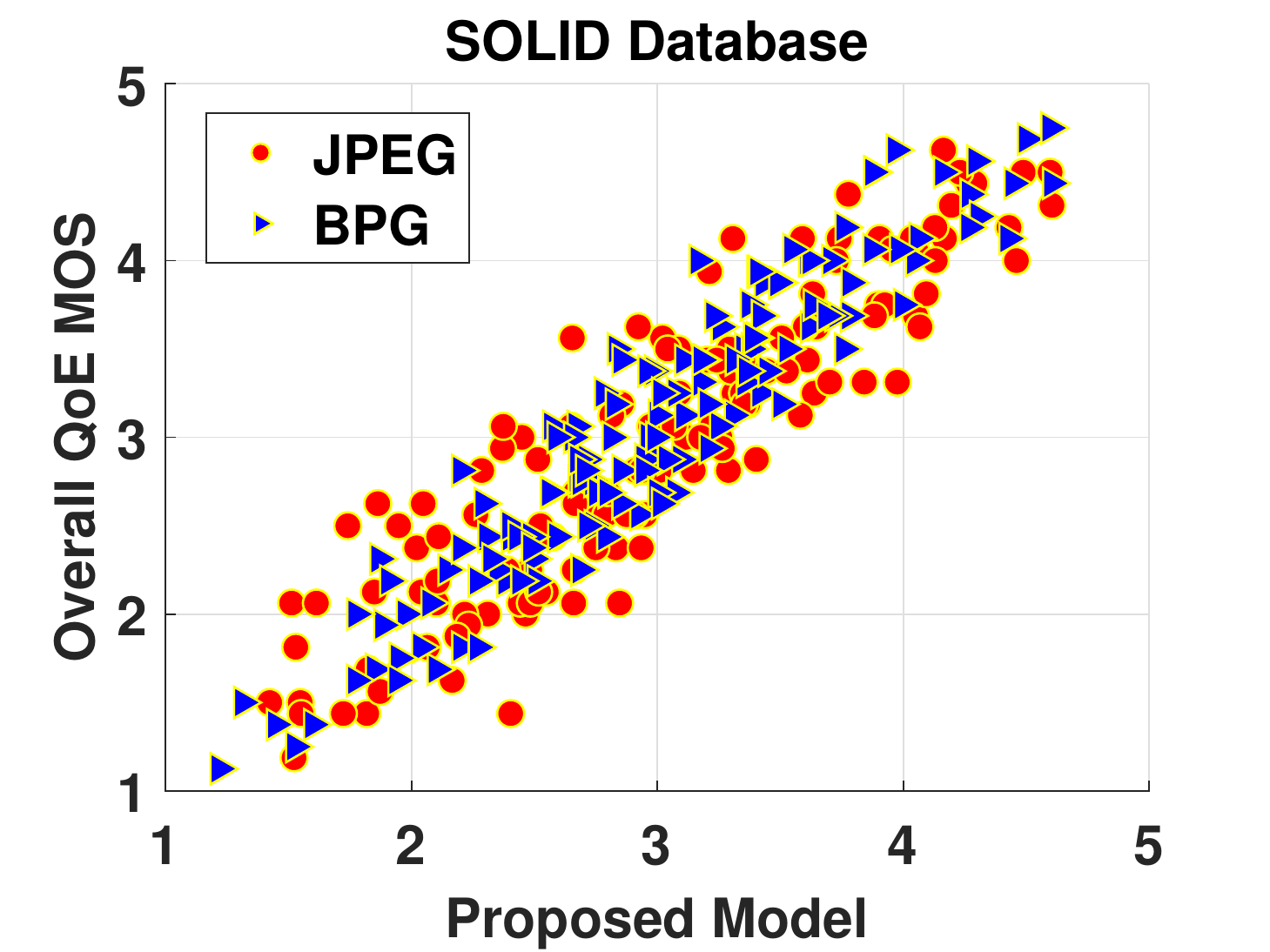}
	}  
	\caption{Scatter plots of MOS against predictions by proposed model on the SOLID database. (a) Image quality, (b) Overall QoE.}
	\label{figure5}
	\vspace{-0.5cm}
\end{figure}

We compare the proposed model with several parametric 2D/3D/360  IQA metrics and the PLCC, SROCC and RMSE performance values are listed in Table \ref{table1}. For 2D metrics such as PSNR, SSIM \cite{wang2004image}, MS-SSIM \cite{wang2003multiscale}, FSIM \cite{zhang2011fsim}, S-PSNR \cite{yu2015framework}, WS-PSNR \cite{sun2017weighted}, CPP-PSNR \cite{zakharchenko2016quality}, the qualities of left and right view images are averaged to obtain the final perceptual score. The simple averaging operation cannot consider the binocular properties of the HVS, thus their correlations with human judgements are not very high. When taking binocular model into consideration, some open source 3D metrics Chen \cite{chen2013full}, W-SSIM \cite{wang2015quality}, W-FSIM \cite{wang2015quality} are tested on the SOLID database, but they also fail to predict the perceptual score of 3D 360 image because the characteristics of omnidirectional images such as FoV and projection deformation are neglected. We further compare our proposed model with stereoscopic omnidirectional image quality evaluator (SOIQE) \cite{chen2019stereoscopic} which is designed for SOIQA. The proposed model shows superior prediction ability to SOIQE. The possible explanation is that our proposed metric adopt a different binocular model and utilize the powerful nonlinearity of SVR to map the qualities of different viewports into a scalar value. Moreover, our proposed method for QoE prediction outperforms the state-of-the-art metrics by a large margin owing to the depth perception related feature extraction. The scatter plots of MOS values versus the predicted scores of the proposed model are drawn in Fig. \ref{figure5}  for image quality and overall QoE to give clear and direct results.

\subsection{Performance Evaluation for Symmetrical/Asymmetrical Distortion}
Symmetrically and asymmetrically distorted images both exist in the SOLID database. The performance on asymmetrical distortion are generally lower than those of symmetrical distortion as shown in Table \ref{table2}, since the binocular fusion, rivalry and suppression may happen in asymmetrically distorted images \cite{chen2018blind}. The proposed method performs best on both symmetrically and asymmetrically distorted images which demonstrates the effectiveness of the binocular model in our algorithm.

\begin{table}[htbp]
\centering
\caption{\textsc{Performance Evaluation for Symmetrically and Asymmetrically Distorted Images on the SOLID Database \cite{xu2018subjective}.}}
\label{table2}
\scalebox{0.75}{
\begin{tabular}{@{}c|cc|cc|cccc@{}}
\toprule
 & \multicolumn{4}{c|}{Symmetrical Distortion} & \multicolumn{4}{c}{Asymmetrical Distortion} \\ \midrule
\multirow{2}{*}{Metric} & \multicolumn{2}{c|}{Image Quality} & \multicolumn{2}{c|}{Overall QoE} & \multicolumn{2}{c|}{Image Quality} & \multicolumn{2}{c}{Overall QoE} \\ \cmidrule(l){2-9} 
 & PLCC & SROCC & PLCC & SROCC & PLCC & \multicolumn{1}{c|}{SROCC} & PLCC & SROCC \\ \midrule
PSNR & 0.791 & 0.789 & 0.705 & 0.707 & 0.394 & \multicolumn{1}{c|}{0.354} & 0.312 & 0.257 \\
SSIM \cite{wang2004image}& 0.944 & 0.902 & 0.840 & 0.813 & 0.821 & \multicolumn{1}{c|}{0.814} & 0.642 & 0.630 \\
MS-SSIM \cite{wang2003multiscale}& 0.869 & 0.836 & 0.761 & 0.736 & 0.631 & \multicolumn{1}{c|}{0.615} & 0.477 & 0.460 \\
FSIM \cite{zhang2011fsim}& 0.930 & 0.890 & 0.833 & 0.805 & 0.853 & \multicolumn{1}{c|}{0.847} & 0.662 & 0.659 \\ \midrule
S-PSNR \cite{yu2015framework}& 0.805 & 0.766 & 0.681 & 0.682 & 0.364 & \multicolumn{1}{c|}{0.313} & 0.361 & 0.222 \\
WS-PSNR \cite{sun2017weighted}& 0.807 & 0.762 & 0.699 & 0.681 & 0.325 & \multicolumn{1}{c|}{0.302} & 0.354 & 0.213 \\
CPP-PSNR \cite{zakharchenko2016quality}& 0.806 & 0.766 & 0.681 & 0.682 & 0.334 & \multicolumn{1}{c|}{0.310} & 0.364 & 0.220 \\ \midrule
Chen \cite{chen2013full}& 0.944 & 0.890 & 0.814 & 0.743 & 0.767 & \multicolumn{1}{c|}{0.700} & 0.522 & 0.434 \\
W-SSIM \cite{wang2015quality}& 0.944 & 0.902 & 0.840 & 0.813 & 0.834 & \multicolumn{1}{c|}{0.832} & 0.643 & 0.638 \\
W-FSIM \cite{wang2015quality}& 0.930 & 0.890 & 0.833 & 0.805 & 0.845 & \multicolumn{1}{c|}{0.842} & 0.652 & 0.658 \\ \midrule
SOIQE \cite{chen2019stereoscopic}& 0.970 & \textbf{0.931} & 0.863 & 0.828 & 0.867 & \multicolumn{1}{c|}{0.866} & 0.718 & 0.717 \\
Proposed & \textbf{0.977} & 0.914 & \textbf{0.962} & \textbf{0.953} & \textbf{0.920} & \multicolumn{1}{c|}{\textbf{0.879}} & \textbf{0.916} & \textbf{0.902} \\ \bottomrule
\end{tabular}}
\vspace{-0.3cm} 
\end{table}

\subsection{Ablation Study}

To verify the effectiveness of each part in our model, we conduct the ablation study as demonstrated in Table \ref{table3}. From this table, we can see that viewport sampling brings huge improvement to the performance. Moreover, weighted averaging according to the binocular model outperforms simply concatenating or averaging the quality scores for both views. In addition, we have tried several methods for computing depth perception features, namely mean, standard deviation and entropy of difference maps. Adopting entropy of the viewport difference map shows the best result for overall QoE prediction.

\begin{table}[htbp]
\centering
\caption{\textsc{Performance Evaluation of Ablation Study.}}
\label{table3}
\begin{threeparttable}
\scalebox{0.82}{
\begin{tabular}{@{}c|ccc|cccc@{}}
\toprule
\multicolumn{4}{c|}{Image Quality} & \multicolumn{4}{c}{Overall QoE} \\ \midrule
Methods & PLCC & SROCC & RMSE & \multicolumn{1}{c|}{Methods} & PLCC & SROCC & RMSE \\ \midrule
w/o VS & 0.889 & 0.885 & 0.464 & \multicolumn{1}{c|}{VS+WQA} & 0.803 & 0.796 & 0.490 \\
VS+QC & 0.929 & 0.917 & 0.365 & \multicolumn{1}{c|}{VS+WQA+DM} & 0.920 & 0.914 & 0.315 \\
VS+QA & 0.921 & 0.910 & 0.382 & \multicolumn{1}{c|}{VS+WQA+DS} & 0.905 & 0.897 & 0.341 \\
VS+WQA & \textbf{0.939} & \textbf{0.928} & \textbf{0.351} & \multicolumn{1}{c|}{VS+WQA+DH} & \textbf{0.935} & \textbf{0.925} & \textbf{0.294} \\ \bottomrule
\end{tabular}}
\begin{tablenotes}
\footnotesize
\item[1] VS denotes viewport sampling.
\item[2] w/o VS denotes the 360-degree stereo image in ERP format.
\item[3] QC, QA and WQA denote viewport quality concatenation, averaging\\ and weighted quality averaging.
\item[4] DM, DS and DH denote the mean, standard deviation and entropy\\ of the viewport difference map.
\end{tablenotes}
\end{threeparttable} 
\vspace{-0.3cm} 
\end{table}

\section{Conclusion}

We propose a novel multi-viewport based stereoscopic omnidirectional image quality assessment metric by considering the FoV and  binocular characteristics of 3D 360 images. It consists of the viewport sampling, binocular quality estimation, depth feature extraction and SVR regression. The experimental results on the public SOLID database prove the effectiveness of our method and it also outperforms several state-of-the-art 2D/3D/360 IQA metrics. In addition, it is shown that the proposed model is able to handle both symmetrical and asymmetrical distortion since we take the binocular properties of the HVS into consideration. Finally, ablation study is conducted to verify the validity of each component in our architecture.

\section*{Acknowledgment}
This work was supported in part by NSFC under Grant 61571413, 61632001.

\bibliographystyle{IEEEtran}
\bibliography{references}

\end{document}